\documentclass[12pt,a4paper]{article}

\usepackage{epsfig}
\usepackage{graphicx}
\usepackage{indentfirst}
\usepackage{multirow}
\usepackage{latexsym}
\usepackage{slashbox}

\date{}

\title{Hexagon Preserving Carbon Nanofoams}

\author{Gotthard Seifert \\
Technische Universitaet Dresden, Bergstr. 66b, 01062 Dresden, Germany \\
and Agnieszka Kuc and Thomas Heine \\
Jacobs University Bremen, Campus Ring 1, 28759 Bremen, Germany}

\begin{document}
\maketitle

\begin{abstract}

Carbon foams are hypothetical carbon allotropes that contain graphite-like (sp$^2$ carbon) segments, connected by sp$^3$ carbon atoms, resulting in porous structures.
 In this work the DFTB (Density Functional based Tight-Binding) method with periodic boundary conditions (PBC) was employed to study energetics, the stability and electronic properties of this unusual class of carbon systems.
Concerning the most stable phases of carbon (graphite and diamond), foams show high structural stability at very low mass density.
The electronic band structure and electronic DOS (density of states) of foams indicate a similar size dependence as carbon nanotubes.
The calculated bulk moduli are in the range between that of graphite (5.5 GPa) and diamond (514 GPa).
These structures may represent novel stable carbon modifications with sp$^2$+sp$^3$ hybridization.
\end{abstract}

\section{Introduction}
\label{sec:1}

Carbon science has been revolutionized by the discovery and synthesis of fulle\-re\-nes\cite{Kroto1985} and nano\-tubes.\cite{Iijima1991}
This revealed that solid carbon, formally believed to exist only as graphite and diamond, was capable to form novel structures.
Carbon atoms can form various types of chemical bonds due to hybridization of $s$ and $p$ orbitals.
Four valence electrons in carbon atoms can form the $sp^{\rm 1}$, $sp^{\rm 2}$ and $sp^{\rm 3}$ bond configurations.
Thus, carbon has the ability to form various allotropes, i.e.\ linear, planar and three-dimensional structures.
Therefore, carbon nanostructures can be classified into seven groups: $sp^{\rm 1}$, $sp^{\rm 2}$, $sp^{\rm 3}$, $sp^{\rm 1}+sp^{\rm 2}$, $sp^{\rm 1}+sp^{\rm 3}$, $sp^{\rm 2}+sp^{\rm 3}$, and $sp^{\rm 1}+sp^{\rm 2}+sp^{\rm 3}$.
At ambient conditions of pressure and temperature, the most stable crystalline carbon allotrope is hexagonal graphite ($sp^{\rm 2}$).
Diamond ($sp^{\rm 3}$), the second stable allotrope of carbon, is at the same conditions nearly as stable as graphite.

Graphite is well known as a moderator in nuclear reactors.
Under irradiation the defined layered structure of graphite changes significantly and many defects are created.
Often occurring defects, vacancies or interstitials can merge into extended defects called dislocation loops or lines.
Aggregation of interstitials cause the interlayer distance expansion due to the formation of new layers.
Vacancy lines, on the other hand, produce the basal contraction.

Recently, considerable advances have been made in the synthesis and theoretical predictions of various carbon nanostructures (for reviews see e.g.\cite{Gogotsi2006, Taylor2009, Benedek2001, Harris1999, Hirsch1999, Jorio2008, Shenderova2002, Gonzalez-Aguilar2007, Ivanovskii2008, Pokropivny2008}).
These nanosystems are often referred to as new carbon allotropes, in addition to the classical allotropes: diamond, graphite and carbyne ($sp^{\rm 1}$).
Among the carbon structures graphenes,\cite{Novoselov2005} onions,\cite{Iijima1980, Kroto1992} diamondoids,\cite{Dahl2003,McIntosh2004} peapods,\cite{Smith1998} scrolls,\cite{Li2005} etc.\ have been synthesized or isolated, and many others were proposed on the basis of theory.\cite{Tomanek2002, Liu1992, Liu1991, Mailhiot1991, McIntosh2004, Hoffmann1983}

In addition to pure \textit{sp}$^2$ or \textit{sp}$^3$ crystals, several experimental and theoretical works focus on the properties of new possible carbon forms, namely those with co-existing \textit{sp}$^2$ and \textit{sp}$^3$ hybridization.\cite{Karfunkel1992, Balaban1994, Umemoto2001, Klett2000, Telling2003, Klett2004}
For instance, the $sp^{\rm 2}+sp^{\rm 3}$ group includes a number of nanoforms with various structures and properties, e.g.\ fullerene polymers, nanotube assemblies, diamond-like crystallites and their combinations.
In addition, diamond-graphite hybrids, vacancies in graphite or carbon foams have been studied.
These investigations include for example pressure-temperature phase transitions of graphite into a cubic diamond structure.\cite{Ribeiro2005}
It was shown that some of the new carbon forms, such as a super-hard carbon phase of C$_{60}$ fullerenes with co-existing \textit{sp}$^2$ and \textit{sp}$^3$ hybridization, appears to have hardnesses higher than that of the (100) and (111) diamond faces.\cite{Blank1995, Serebryanaya2001, Talyzin2002}

A very interesting family of carbon nanostructures are so-called Carbon Nano\-foams.
Carbon nanofoams can be thought of as graphite-based materials with significantly enlarged interlayer distance.
In carbon nanofoams some of the \textit{sp}$^2$ carbon atoms are transformed into \textit{sp}$^3$ carbon atoms, which allows diamond-like fragments to co-exist with graphene fragments.
Dislocations in graphite\cite{Heggie2006, Suarez2007} lead to other type of foam-based carbon nanostructures, where the graphene planarity is lost, but two types of systems can be found: those with pure $sp^{\rm 2}$ hybridization or those with co-existing $sp^{\rm 2}+sp^{\rm 3}$ hybridization of atoms.

Single-walled carbon foams might be formed by hierarchical self-assembly processes from layered graphite\cite{Strobel2006} or by a cold compression of carbon nano\-tubes.\cite{Wang2004}
The experiment of Wang et al.,\cite{Wang2004} where a sample of carbon nano\-tubes has been coldly compressed in a diamond anvil cell, shows the transformation into whatis believed to be a novel carbon allotrope.
It was discussed on the basis of theoretical calculations\cite{Bucknum2006} that the new form of carbon obtained by Wang et al.\cite{Wang2004} can be described as a carbon foam structure.
The results indicate that kinetically stabilized products, such as low-density carbon foam materials, may have possibly been formed in this experiment.
It was also shown that the new crystalline material is a hard carbon phase with a high (calculated) bulk modulus.\cite{Bucknum2006}

In 1987, Vanvechten et al\cite{Vanvechten1987}.\ showed that dense packing of C$_{11}$ clusters consisting of three five-fold rings in a condensed phase can result in a foam-like system.
The structure of carbon foams was, however, proposed only later by Karfunkel et al\cite{Karfunkel1992}.\ and Balaban et al.\cite{Balaban1994}
Recently, experimental\cite{Klett2004, Klett2000, Wang2004} and theoretical\cite{Umemoto2001, Park2000, Bucknum2006} investigations have shown that carbon-foam-like materials can be formed by rather simple syntheses.
For example, the mesophase pitch precursor is molten at high temperatures resulting in so-called graphitic foams.
Although these systems are no single-wall carbon foams, as they are discussed in this work, it is possible to achieve similar structures experimentally.

Recently, Braga et al.\cite{Braga2007} have shown, using molecular dynamics simulations, that after applying compression small diameter carbon nanotubes (CNTs) undergo polymerization.
The resulting structures belong to the $sp^{\rm 2}+sp^{\rm 3}$ group and are similar to carbon foams.
Nanotube-derived foams are also discussed by Ding et al.\cite{Ding2007}
The new carbon forms are designed theoretically based on the welding technique using cross-linkers.
Such linkers create covalently bound nanotube array with high porosity and well-separated CNTs.
Cross-linkage of nanotubes was also successfully performed in experiments.\cite{Leonard2009}

Another interesting carbon allotrope of the $sp^{\rm 2}+sp^{\rm 3}$ group is the so-called glitter.\cite{Bucknum1994}
It can be thought of  1,4-cyclohexadienoid motifs connected into a 3D structure, forming a 3D network of channels, in contrast to carbon nanofoams, which can be either 1D or 2D porous structures.
Glitter is a hypothetical carbon allotrope but can be viewed as a plausible model of n-diamond.
This system is predicted to be as hard as diamond.
The cross section of glitter reminds one of the possible carbon foam structures, therefore it is discussed in this work as well.

\begin{figure}[ht]
\includegraphics[scale=.75]{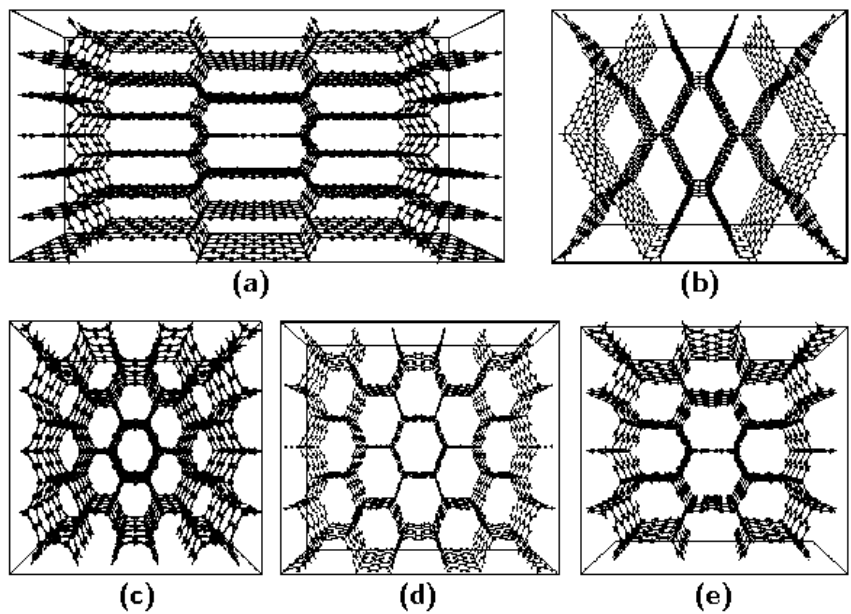}
\caption{Exemplary carbon foam structures: (5,1)- a, (1,5) b, (1,1) -c zig-zag carbon foams: (3,3) d, (3,2) - e armchair carbon foams (see text for the nomenclature).}
\label{fig:1}
\end{figure}

\begin{figure}[ht]
\includegraphics[scale=.65]{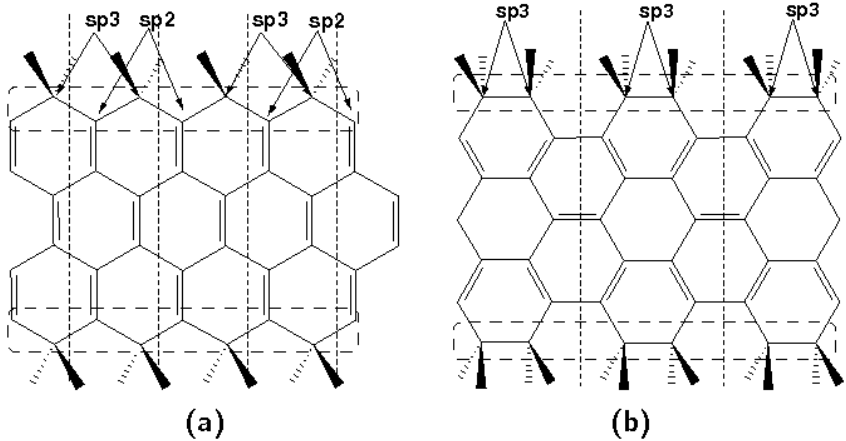}
\caption{The boundary atoms of zig-zag (a) and armchair (b) carbon foams. The unit cells are indicated by vertical lines.\cite{Kuc2006}}
\label{fig:2}
\end{figure}

In this chapter, single-walled carbon foams\cite{Kuc2006} [see Fig.~\ref{fig:1}], defected graphite and glitter are discussed.
Due to the pattern of open edges, two types of such structures can be built, so-called armchair and zig-zag foams, in analogy to the nomenclature of carbon nano\-tubes.
The pore size is defined by a pair of integer numbers ($N$,$M$), which indicate the number of hexagonal units between the junctions.
The junctions consist of the \textit{sp}$^3$ boundary-atom chains [cf.\ Fig.~\ref{fig:2}].
The two numbers are necessary to distinguish between the possible symmetric carbon foams of size $N=M$ and asymmetric ones with $N\neq M$.
This nomenclature will be use in the following.

\section{General Considerations and Computational Methods}
\label{subsec:1}

The binding energies of the carbon foams can be considered as follows:
The elementary cell consists of $n_x$ carbon atoms at the boundaries (junctions) and $n_i$ atoms inside the graphene fragments (stripes).
Thus, the energy of a carbon foam can be written as
\begin{equation}
\label{eq:1}
E_{\rm bind}=n_i\epsilon_{\infty}+n_x\epsilon_x,
\end{equation}
where $\epsilon_{\infty}$ is the binding energy of the infinite graphene layer, whereas $\epsilon_x$ describes the binding energy of the boundary atoms in the unitcell.
Since the total number of atoms $n=n_i+n_x$, the energy can be expressed as follows:
\begin{equation}
\label{eq:2}
E_{\rm bind}=n\epsilon_{\infty}-n_x\epsilon_{\infty}+n_x\epsilon_x.
\end{equation}
Defining $\Delta\epsilon=\epsilon_x-\epsilon_{\infty}$, the energy per atom becomes:
\begin{equation}
\label{eq:3}
\frac{E_{\rm bind}}{n}=\epsilon_{\infty}+\frac{n_x\Delta\epsilon}{n},
\end{equation}
where the number of boundary atoms $n_x$ per unit cell is hold constant at $n_x=8$.

\begin{figure}[ht]
\includegraphics[scale=.75]{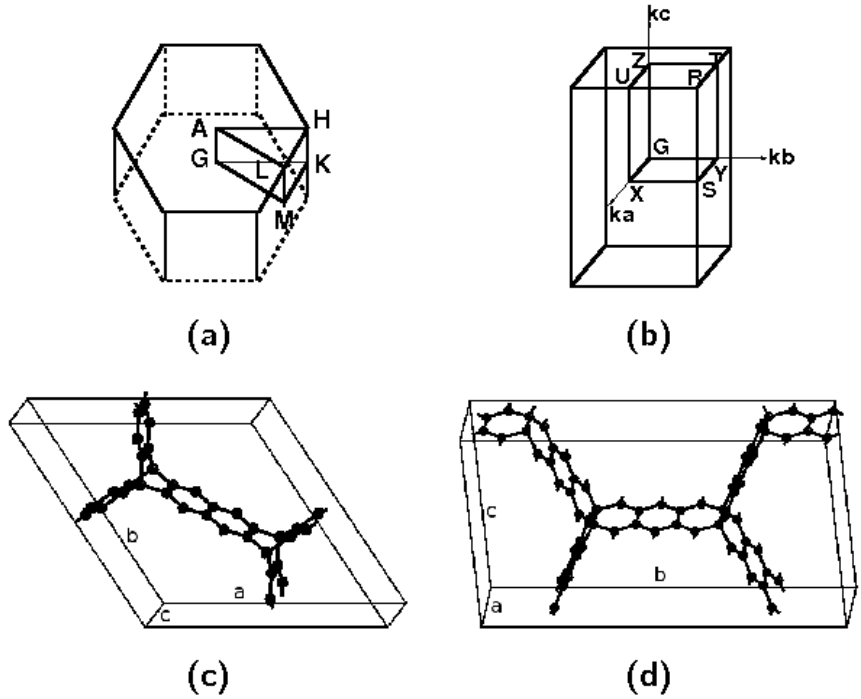}
\caption{The hexagonal (a) and orthorhombic (b) Brillouin zones for the respective elementary unit cells (c and d) of the (3,3) zig-zag carbon foam.\cite{Kuc2006}}
\label{fig:3}
\end{figure}

The detailed and quantitative studies of the stability, electronic and mechanical properties of dif\-fer\-ent carbon foams were performed using the Density Functional based Tight-Binding\cite{Seifert1996} (DFTB) method.
Periodic boundary conditions were used to calculate the infinite crystalline solid state.
The conjugate-gradient scheme was chosen for the geometry optimization.
The number of \textbf{k}-points was determined by reaching convergence for the total energy as a function of \textbf{k}-points according to the scheme proposed by Monkhorst and Pack.\cite{Monkhorst1976}
Band structures were computed along lines between high symmetry points of the Brillouin zone.
The first Brillouin zones with the highly symmetric points for hexagonal and orthorhombic unit cells [Fig.~\ref{fig:3}a and Fig.~\ref{fig:3}b] are shown in Fig.~\ref{fig:3}c and Fig.~\ref{fig:3}d.

The mechanical properties were studied in detail by estimation of the bulk and shear moduli.
The elastic constants (stiffness) $c_{ij}$ were calculated using the finite-dif\-fer\-ence scheme (the derivatives of the total energy with strain $\epsilon_{i,j}$) as
\begin{equation}
\label{eq:4}
\frac{\partial}{\partial\epsilon_i}\times\left(\frac{\partial E}{\partial\epsilon_j}\right)=c_{ij}.
\end{equation}
The matrix of constants was further used to obtain the bulk modulus
\begin{equation}
\label{eq:5}
\mathbf{B}=\frac{1}{9}\left[c_{11}+c_{22}+c_{33}+2\left(c_{12}+c_{13}+c_{23}\right)\right]
\end{equation}
for orthorhombic lattices or
\begin{equation}
\label{eq:6}
\mathbf{B}=\frac{\Delta c_{33}+2c_{13}}{\Delta + 2}
\end{equation}
for hexagonal unit cells, where
\begin{equation}
\label{eq:7}
\Delta=\frac{c_{11}+c_{12}-2c_{13}}{c_{33}-c_{13}}.
\end{equation}
The shear modulus \textbf{G} can be calculated according to:
\begin{equation}
\label{eq:8}
\mathbf{G}=\frac{1}{15}\left[\left(c_{11}+c_{22}+c_{33}-c_{12}-c_{13}-c_{23}\right)+3\left(c_{44}+c_{55}+c_{66}\right)\right].
\end{equation}

\section{Structure}
\label{subsec:21}

\begin{figure}[ht]
\includegraphics[scale=.15]{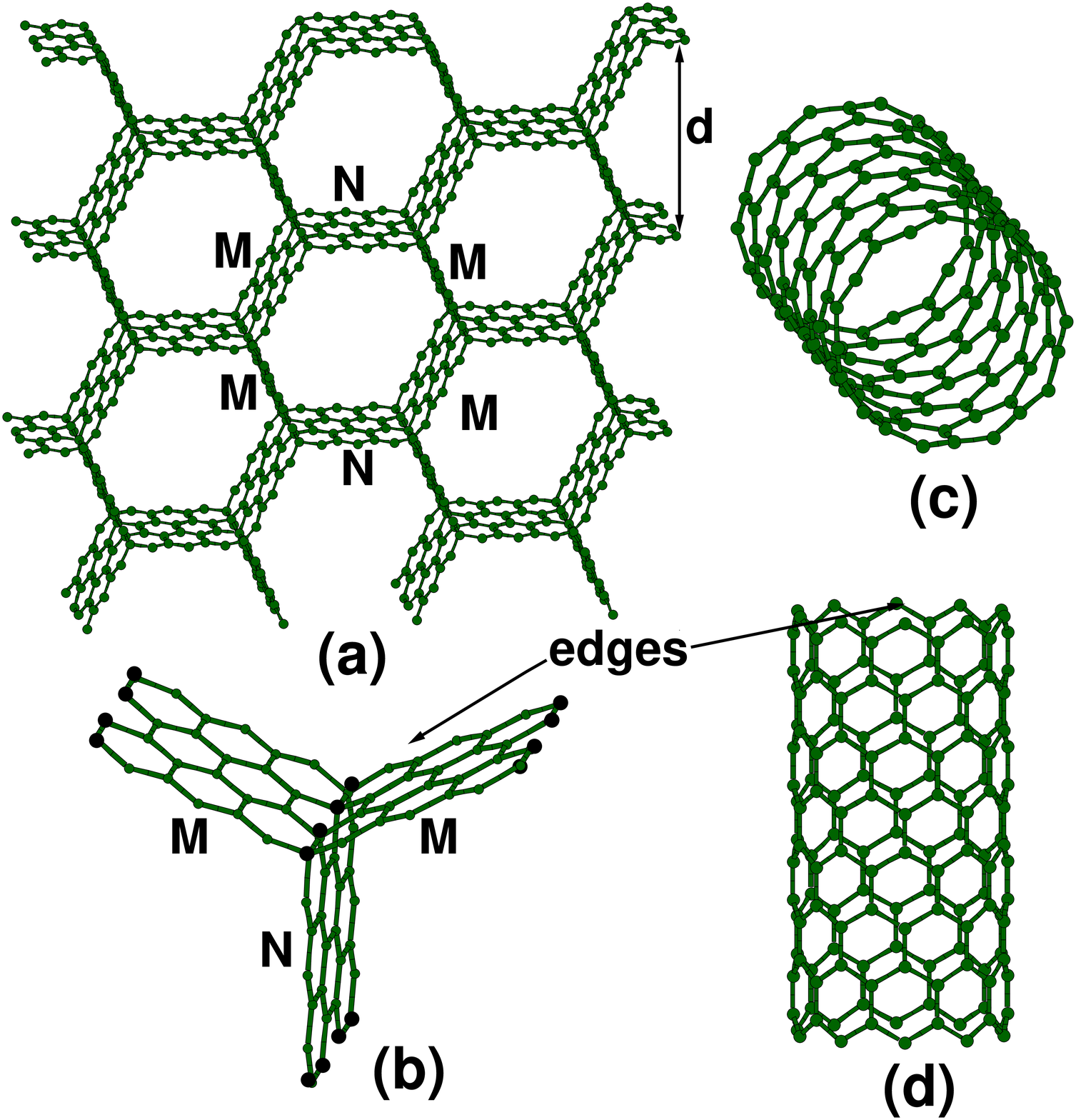}
\includegraphics[scale=.15]{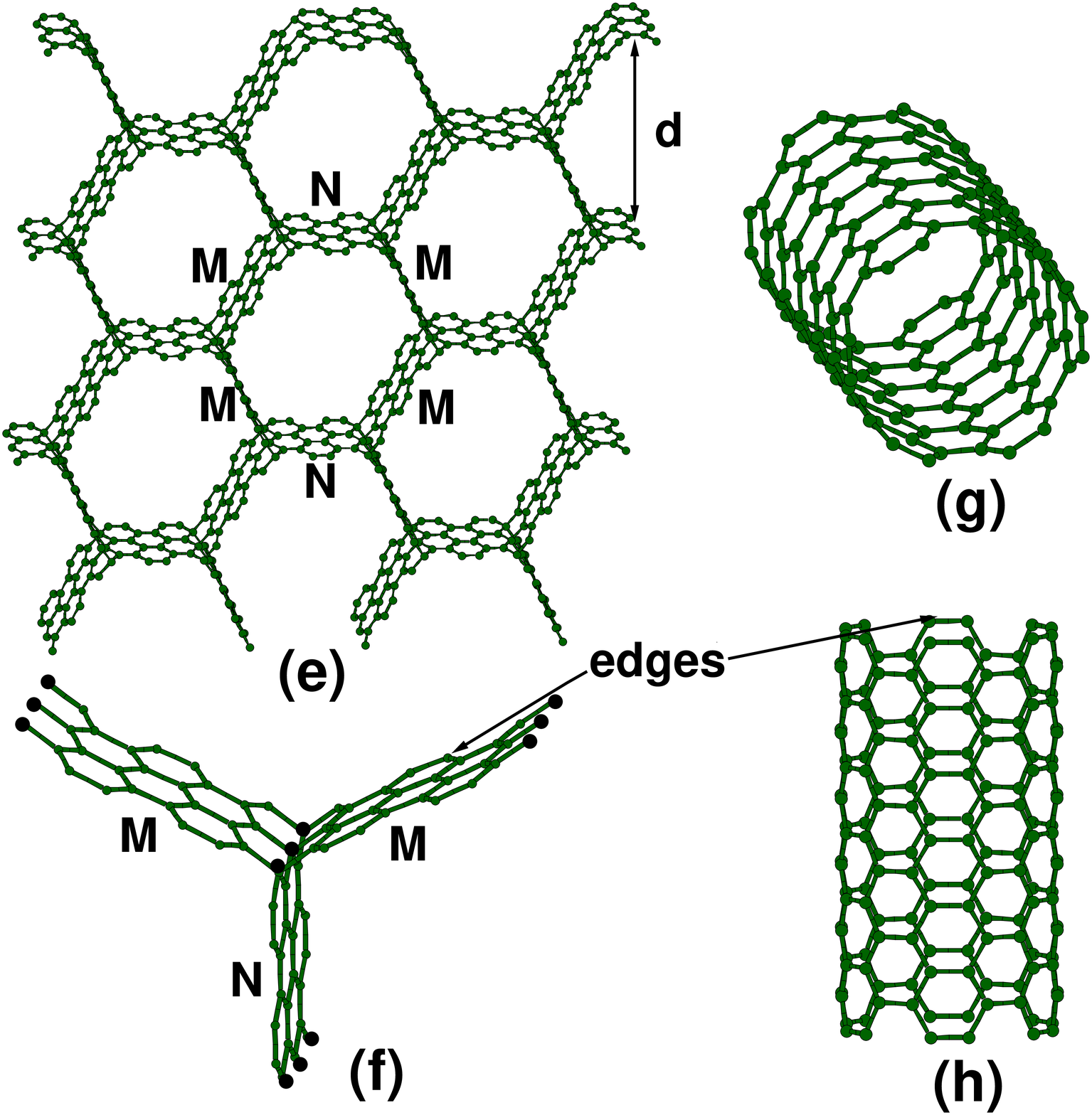}
\caption{(a) The structure of the (3,3) zig-zag carbon foam and (b) its bonding configuration at the junction in an orthorhombic lattice compared to the(10, 0) zig-zag carbon nano\-tube (c, d). (e) The structure of the (3,5) armchair carbon foam and (f) its bonding configuration at the junction in an orthorhombic latticecompared to the (5, 5) armchair carbon nano\-tube (g, h). Only three hexagonal units in $a$ direction and two unit cells in $b$ and $c$ directions are shown for visual clarity. The black circles represent the linear chain of the \textit{sp}$^3$ hybridized atoms.}
\label{fig:4}
\end{figure}

As described above, the Carbon foams are three-dimensional porous structures that contain both \textit{sp}$^2$ and \textit{sp}$^3$ hybridized atoms [see Fig.~\ref{fig:1}].
They can be thought of as constructed from graphene planes interconnected rigidly with one another at 120$^\circ$, forming a linear chain of \textit{sp}$^3$ bonded atoms along the junction [see Fi\-gures~\ref{fig:4}].
At these junctions, always three graphene layers meet [c.f\ Fi\-gures~\ref{fig:4} (c) and (f)] resulting in the honey-comb-shaped cross-section (CS) of the foam.
If the graphitic segments are connected to each other at dif\-fer\-ent angles than 120$^\circ$, the systems with non-hexagonal CS are obtained.
The present discussion was restricted to foams that preserve six-fold rings and hexagonal cross-sections.\cite{Kuc2006}

Carbon foams discussed here can also be considered as AA-stacked gra\-phite structures with a significantly increased interlayer distance.
Thus, $N$ describes the length of graphene fragments linked together by $M$ graphene stripes.
In other words, the $N$-stripes are functionalized by $M$-fragments.
In this way, $M$ determines the interlayer spacing as $d = \sqrt{3} M$.
The ABAB-stacked forms are possible as well, as e.g.\ in the case of defected graphite [Sect.~\ref{subsec:5}].
The \textit{sp}$^2$ carbon atoms, which are transformed into \textit{sp}$^3$ atoms, make rigid interconnections between the graphene layers.

In Fig.~\ref{fig:4}a and Fig.~\ref{fig:4}e perspective views of the (3,3) zig-zag and the (3,5) armchair carbon foams are given, in comparison to the (10,0) zig-zag [Fig.~\ref{fig:4}c and Fig.~\ref{fig:4}d] and (5,5) armchair [Fig.~\ref{fig:4}g and Fig.~\ref{fig:4}h] carbon nano\-tubes.

For a given pore size, the elementary unit cells of zig-zag and armchair carbon foams dif\-fer in the number of atoms ($n$).
As an example, the (2,2) carbon foam consists of 44 and 20 carbon atoms for the zig-zag and the armchair foam, respectively.
All structures can be represented in orthorhombic 3D carbon networks.
Moreover, the zig-zag foams, with $N=M$, can be described within hexagonal lattices as well.

The interlayer distances \textit{d} [Fig.~\ref{fig:4}] vary in the range of 4.7~\AA\ [(1,1) zig-zag foam] and 32.3~\AA\ [(9,9) armchair foam].
The orthorhombic unit cell length in $a$ direction can be as short as 2.46~\AA\ and 4.27~\AA\ for armchair and zig-zag carbon foams, respectively.
The pore size is determined by the unit cell parameters $b$ and $c$ [see Fig.~\ref{fig:3} for definition].
Keeping $a$ at its minimum and reducing the width in the $b$ and $c$ directions to zero ($N$,$M=0$), the structure reduces to a network of fourfold coordinated carbon atoms, namely that of cubic diamond (from armchair carbon foams) or hexagonal diamond (isodiamond; from zig-zag carbon foams).\cite{Balaban1994, Wang2004a}
On the other hand, increasing the system in $a$ and $b$ directions gives, in the $a, b \longrightarrow \infty$ limit, the structure of an isolated graphene layer.

The structures of zig-zag and armchair foams also dif\-fer in the types of connections (bonds): three types of covalent bonds, \textit{sp}$^2-$\textit{sp}$^2$, \textit{sp}$^2-$\textit{sp}$^3$, and \textit{sp}$^3-$\textit{sp}$^3$, can be found in the zig-zag arrangement [cf.\ Table~\ref{tab:1}], whereas armchair foams have two kinds of \textit{sp}$^2-$\textit{sp}$^2$ bonds [single and double; cf.\ Tabel~\ref{tab:2}] and \textit{sp}$^2-$\textit{sp}$^3$ bonds (there are no direct \textit{sp}$^3-$\textit{sp}$^3$ connections along the $a$ axis).
Comparing the geometries of carbon foams with the corresponding data for graphite and diamond [Tabel~\ref{tab:1}], one can find that the bond lengths are in between the values for both carbon allotropes.
The \textit{sp}$^3-$\textit{sp}$^3$ bonds at the junctions are only slightly distorted from the ideal tetrahedral bonds in diamond.
Moreover, the angles at the threefold and fourfold coordinated atoms are the same as those in graphite and diamond, respectively.
The \textit{sp}$^2-$\textit{sp}$^2$ bond lengths in the zig-zag foams are very close to the bond lengths in graphite, whereas in armchair foams 'single' and 'double' bonds between \textit{sp}$^2$ carbon atoms exist [cf.\ Tab.~\ref{tab:2}].
The dif\-fer\-ence between 'single' and 'double' bonds decreases with increasing size, approaching the value of graphite.
\begin{table}
\caption{The geometry parameters of the symmetric zig-zag carbon foams compared to the calculated and the experimental data for graphite and diamond (values in parenthesis).\cite{Kuc2006}}
\label{tab:1}
\begin{tabular}{p{1.5cm}p{2.3cm}p{2.3cm}p{2.3cm}p{2.3cm}}
\hline\noalign{\smallskip}
\textbf{Structure} & \multicolumn{2}{c}{{Bond length (\AA)}} & \multicolumn{2}{c}{{Bond angle ($^\circ$)}} \\
\hline\noalign{\smallskip}
\textbf{graphite} & \textit{sp}$^2-$\textit{sp}$^2$ & 1.420 (1.421) & \textit{sp}$^2-$\textit{sp}$^2-$\textit{sp}$^2$ & 120.11 (120.00)\\
\textbf{diamond} &  \textit{sp}$^3-$\textit{sp}$^3$ & 1.541 (1.545) & \textit{sp}$^3-$\textit{sp}$^3-$\textit{sp}$^3$ & 109.47 (109.47) \\
\textbf{(1,1)} & \textit{sp}$^2-$\textit{sp}$^2$ & 1.359 & \textit{sp}$^2-$\textit{sp}$^2-$\textit{sp}$^2$ & 120.00 \\
& \textit{sp}$^2-$\textit{sp}$^3$ & 1.536 & \textit{sp}$^3-$\textit{sp}$^2-$\textit{sp}$^3$ & 109.11 \\
& \textit{sp}$^3-$\textit{sp}$^3$ & 1.557 &  \\
\textbf{(2,2)} & \textit{sp}$^2-$\textit{sp}$^2$ & 1.418 & \textit{sp}$^2-$\textit{sp}$^2-$\textit{sp}$^2$ & 120.00 \\
& \textit{sp}$^2-$\textit{sp}$^3$ & 1.534 & \textit{sp}$^3-$\textit{sp}$^2-$\textit{sp}$^3$ & 109.16 \\
& \textit{sp}$^3-$\textit{sp}$^3$ & 1.585 &  \\
\textbf{(3,3)} & \textit{sp}$^2-$\textit{sp}$^2$ & 1.420 & \textit{sp}$^2-$\textit{sp}$^2-$\textit{sp}$^2$ & 120.00 \\
& \textit{sp}$^2-$\textit{sp}$^3$ & 1.530 & \textit{sp}$^3-$\textit{sp}$^2-$\textit{sp}$^3$ & 109.17\\
& \textit{sp}$^3-$\textit{sp}$^3$ & 1.567 &  \\
\textbf{(4,4)} & \textit{sp}$^2-$\textit{sp}$^2$ & 1.424 & \textit{sp}$^2-$\textit{sp}$^2-$\textit{sp}$^2$ & 120.00 \\
& \textit{sp}$^2-$\textit{sp}$^3$ & 1.531 & \textit{sp}$^3-$\textit{sp}$^2-$\textit{sp}$^3$ & 109.16\\
& \textit{sp}$^3-$\textit{sp}$^3$ & 1.576 &  \\
\textbf{(5,5)} & \textit{sp}$^2-$\textit{sp}$^2$ & 1.425 & \textit{sp}$^2-$\textit{sp}$^2-$\textit{sp}$^2$ & 120.00 \\
& \textit{sp}$^2-$\textit{sp}$^3$ & 1.532 & \textit{sp}$^3-$\textit{sp}$^2-$\textit{sp}$^3$ & 109.17 \\
& \textit{sp}$^3-$\textit{sp}$^3$ & 1.575 &  \\
\hline\noalign{\smallskip}
\end{tabular}
\end{table}

\begin{table}
\caption{The geometry parameters of the symmetric armchair car\-bon foams.\cite{Kuc2006}}
\label{tab:2}
\begin{tabular}{p{1.5cm}p{2.3cm}p{2.3cm}p{2.3cm}p{2.3cm}}
\hline\noalign{\smallskip}
\textbf{Structure} & \multicolumn{2}{c}{{Bond length (\AA)}} & \multicolumn{2}{c}{{Bond angle ($^\circ$)}} \\
\hline\noalign{\smallskip}
\textbf{(2,2)} & \textit{sp}$^2-$\textit{sp}$^2$ & 1.352, 1.448 & \textit{sp}$^2-$\textit{sp}$^2-$\textit{sp}$^2$ & 120.87 \\
&  \textit{sp}$^2-$\textit{sp}$^3$ & 1.523 & \textit{sp}$^3-$\textit{sp}$^2-$\textit{sp}$^3$ & 110.46\\
\textbf{(3,3)} & \textit{sp}$^2-$\textit{sp}$^2$ & 1.372, 1.442 & \textit{sp}$^2-$\textit{sp}$^2-$\textit{sp}$^2$ & 120.63 \\
& \textit{sp}$^2-$\textit{sp}$^3$ & 1.518 & \textit{sp}$^3-$\textit{sp}$^2-$\textit{sp}$^3$ & 110.07\\
\textbf{(4,4)} & \textit{sp}$^2-$\textit{sp}$^2$ & 1.384, 1.436 & \textit{sp}$^2-$\textit{sp}$^2-$\textit{sp}$^2$ & 120.41\\
& \textit{sp}$^2-$\textit{sp}$^3$ & 1.515 & \textit{sp}$^3-$\textit{sp}$^2-$\textit{sp}$^3$ & 109.23\\
\textbf{(5,5)} & \textit{sp}$^2-$\textit{sp}$^2$ & 1.394,1.443 & \textit{sp}$^2-$\textit{sp}$^2-$\textit{sp}$^2$ & 120.27\\
& \textit{sp}$^2-$\textit{sp}$^3$ & 1.515 & \textit{sp}$^3-$\textit{sp}$^2-$\textit{sp}$^3$ & 108.81\\
\hline\noalign{\smallskip}
\end{tabular}
\end{table}

\begin{figure}[ht]
\includegraphics[scale=.45]{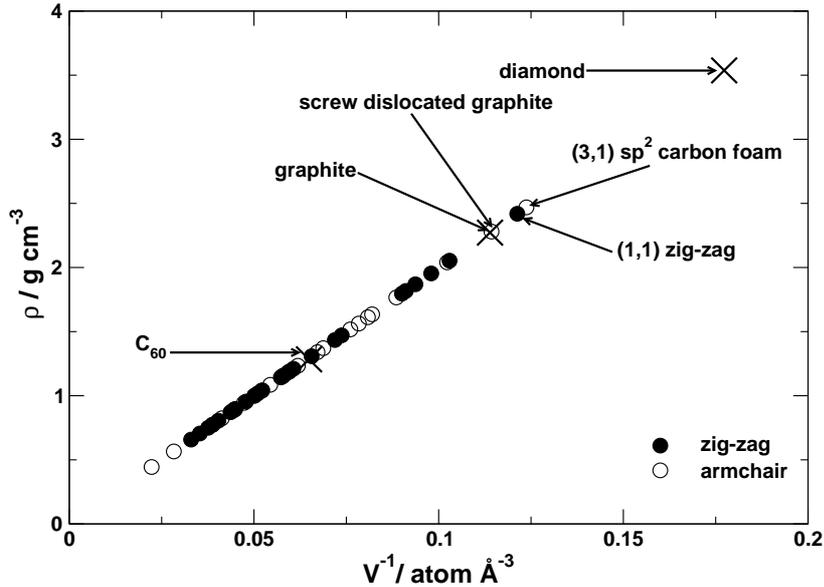}
\caption{The mass densities versus $V^{-1}$ ($V$-the atomic volume) of carbon foams compared to C$_{\rm {60}}$, graphite and diamond.\cite{Kuc2006}}
\label{fig:5}
\end{figure}

Furthermore, the results indicate that the optimized unit cell sizes of carbon foams correspond to mass densities smaller than that of graphite ($\rho=2.27$~g~cm$^{-3}$) and diamond ($\rho=3.55$~g~cm$^{-3}$) [cf.\ Fig.~\ref{fig:5}].
The only exception was found for the (1,1) zig-zag foam with a mass density of $\rho=2.42$~g~cm$^{-3}$.

\begin{figure}[ht]
\includegraphics[scale=.22]{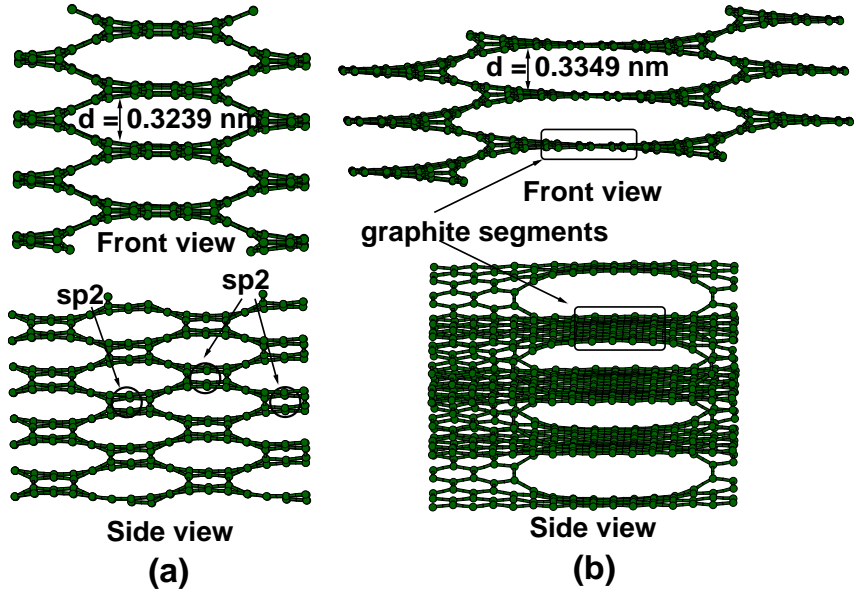}
\caption{The structure of \textit{sp}$^2$ carbon foams: (a) the (3,1) armchair carbon foam and (b) screw dislocated graphite.}
\label{fig:6}
\end{figure}

The smallest armchair foams ($N$,$1$) with an initial distance between the graphitic segments smaller than the van der Waals interlayer distance of graphite become very interesting systems after a full geometry optimization [Fig.~\ref{fig:6}].
During the optimization, the \textit{sp}$^3$ hybridized atoms of these structures open one of the four bonds and bind strongly to three neighbors only.
This results in a porous system, built of \textit{sp}$^2$ carbon atoms (in the following these forms are called '\textit{sp}$^2$ carbon foams').
To obtain such structures a larger unit cell in $a$ direction is required (at least three times the original lattice) to allow the breaking of \textit{sp}$^3-$\textit{sp}$^3$ connections.
Otherwise the optimization leads to a typical diamond system.
As an example the (3,1) \textit{sp}$^2$ armchair foam is shown in Fig.~\ref{fig:6}a.
Its distance $d=3.24$~\AA\ between the graphitic fragments as well as the bond lengths are similar to those of layered graphite, although the arrangement of the atoms isdif\-fer\-ent.
The mass density ($\rho=2.47$~g~cm$^{-3}$) of this foam is about the same as for the (1,1) zig-zag carbon foam.
Analogous \textit{sp}$^2$ foams were also discussed by Umemoto et al.\cite{Umemoto2001} with similar conclusions.

Similar to the \textit{sp}$^2$ foams, carbon foams can be constructed by screw twisting of graphite layers.
They can be related to the structure of 'Screw Dislocated Graphite (SDG)\cite{Suarez2007}' that is shown in Fig.~\ref{fig:6}b.
This type of carbon foams will be called 'screw dislocated graphite' or SDC in the following.
SDC has the same interlayer distance and bond lengths as layered graphite.
However, there are covalent bonds present in $c$ direction that connect the neighboring graphene fragments parallel to the $ab$ planes.
By construction, these bonds are formed by providing atoms within the graphene layers with additional neighbors in $c$ direction, locally removing planarity without changing the hybridization of the carbon atoms.
The system - shown in Fig.~\ref{fig:6}b - corresponds to the SDG (7,1) armchair carbon foam.

Both types of \textit{sp}$^2$ foams have dif\-fer\-ent kinds of nano\-pores, but both form two-dimensional interconnected channels between the pores: the \textit{sp}$^2$carbon foams have direct connections between pores along the $b$ direction [Fig.~\ref{fig:6}a bottom], while SDG connections are rather twisted [Fig.~\ref{fig:6}b bottom].
On the other hand, the \textit{sp}$^2+$\textit{sp}$^3$ carbon foams have closed nano\-pores (one-dimensional channels), similar to nano\-tubes.

\section{Energetics and Mechanical Stability}
\label{subsec:3}

\begin{figure}[ht]
\includegraphics[scale=.45]{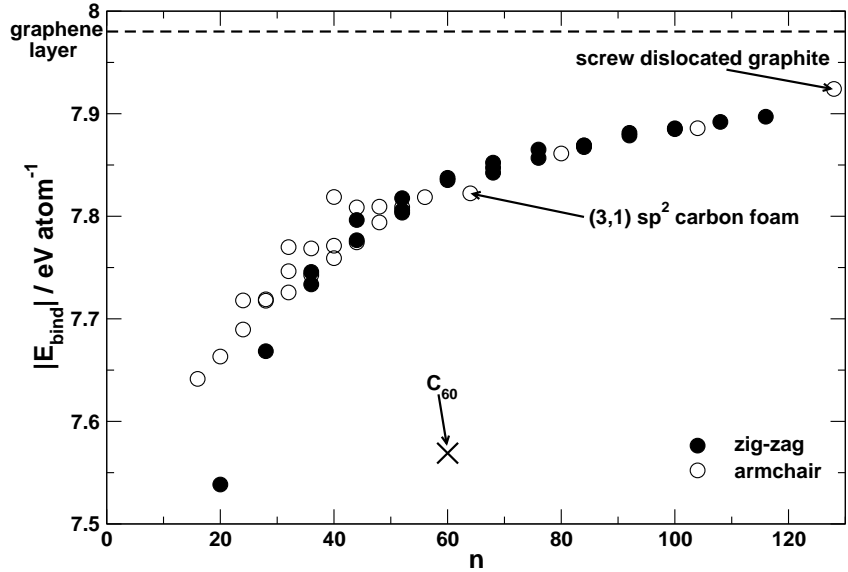}
\caption{The binding energy of the investigated carbon foams as a function of $n$ ($n$-the number of atoms per unit cell). The corresponding energy of asingle graphene layer is given as a reference (dashed line).\cite{Kuc2006}}
\label{fig:7}
\end{figure}

\begin{figure}[ht]
\includegraphics[scale=.45]{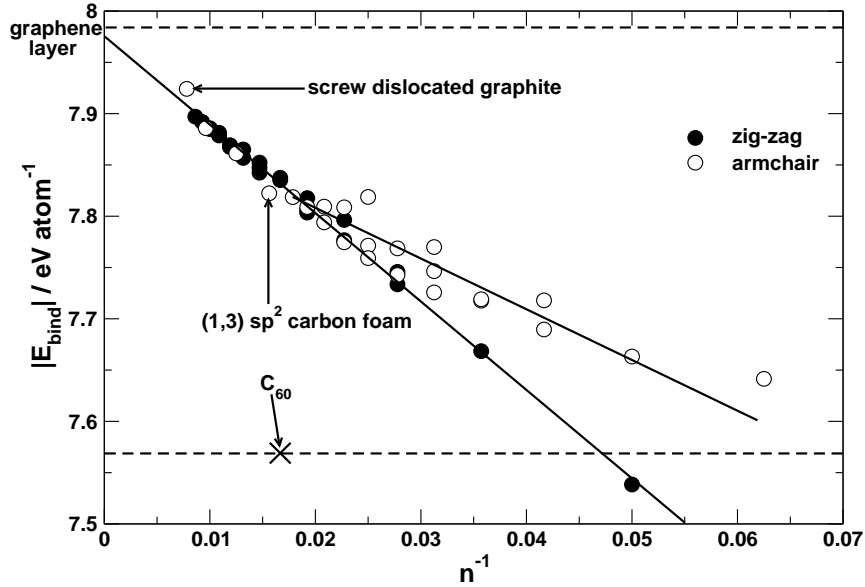}
\caption{The binding energy of carbon foams as a function of $n^{-1}$.The corresponding energy of a single graphene layer is given as a reference (dashed line).\cite{Kuc2006}}
\label{fig:8}
\end{figure}

The energetic and mechanical stability of dif\-fer\-ent carbon foams was studied following the discussion in Sect.~\ref{subsec:1}.
The binding energy (per atom) as a function of $n$ ($n$ is the number of atoms per unit cell) is shown in Fig.~\ref{fig:7}.
The energy of carbon foams asymptotically approaches the binding energy of a graphene layer.
According to the proposed model consideration [Sect.~\ref{subsec:1}] the cohesive energy should follow a linear trend with respect to $n^{-1}$.
Figure~\ref{fig:8} shows that the energy of the investigated carbon foams indeed increases nearly linearly with $n^{-1}$, as expected from Eq.~\ref{eq:3}.
The deviations from the linearity [cf.\ Fig.~\ref{fig:8}] may be explained by the dif\-fer\-ences in the types of boundary atoms ($n_x$) for a given type of structures.
In zig-zag systems there are 8 \textit{sp}$^3$ junction atoms per elementary unit cell, while 4 \textit{sp}$^2$ and  4 \textit{sp}$^3$ carbon atoms (per unit cell) occurat the junctions in armchair structures [cf.\ Fig.~\ref{fig:2}].
The \textit{sp}$^2$ carbon atoms in zig-zag foams form graphene stripes with a fully delocalized $\pi$-electron system.
In armchair foams, the $\pi$-electron delocalization is distorted by the $\pi$ bonds at the \textit{sp}$^2$ atoms at the boundary of graphene-like stripes.
This has obviously a stronger influence in smaller structures.
For larger systems the bonding behavior of all boundary atoms becomes very similar, i.e.\ all structures have nearly the same $n^{-1}$ size dependence.
This is also confirmed by the observation, that the \textit{sp}$^2-$\textit{sp}$^2$ bond lengths in armchair systems depend on the size:
The single and double bond lengths become similar with increasing the system size [Tab.~\ref{tab:2}].
They slowly approach the values of bond lengths in graphite.

The calculations indicate that carbon foams are quite stable systems compared to the other well-known carbon allotropes.
The origin of their favorable stability is the fact that the carbon foams discussed here do not contain bents, but only straight graphitic planes, in contrast to fullerenes and nano\-tubes that are purely \textit{sp}$^2$ bonded but exhibit curved graphitic fragments.
The largest studied carbon foams [(5,5) zig-zag ($n=116$) and (7,7) armchair ($n=104$)] are almost as stable as graphite and diamond.
Their cohesive energies were found to be smaller by only $\sim$0.09~eV~atom$^{-1}$ than that of a graphene layer ($E_{\rm bind}=7.986$~eV~atom$^{-1}$).
The least stable (1,1) zig-zag carbon foam ($n=20$) with an energy of 7.538~eV~atom$^{-1}$ is as stable as the (5,5) armchair carbon nano\-tube ($E_{\rm bind}=7.539$~eV~atom$^{-1}$) and similarly stable to the (10,0) zig-zag carbon nano\-tube ($E_{\rm bind}=7.501$~eV~atom$^{-1}$).
Furthermore, it was found that except for the (1,1) zig-zag carbon foam, all structures are more stable than the isolated C$_{60}$ cage ($E_{\rm bind}=7.569$~eV~atom$^{-1}$) by at least $\sim$0.08~eV~atom$^{-1}$.
The \textit{sp}$^2$ carbon foams are very stable, as well.
As an example, the (3,1) foam is less stable by $\sim$0.16~eV~atom$^{-1}$ than the most stable carbon allotropes.
The stability of SDG (7.924~eV~atom$^{-1}$) and graphite/diamond is about the same.
These results suggest that carbon foams can be stable once they have been formed.

The mechanical properties were studied by estimation of the bulk (\textbf{B}) and shear (\textbf{G}) moduli according to Eq.~\ref{eq:5} and Eq.~\ref{eq:8}.
The results [Tab.~\ref{tab:3} and Tab.~\ref{tab:4}] show that with increasing size of the pores ($N$ and/or $M$) the bulk modulus decreases.
The smallest \textbf{B} belongs to the (1,5) zig-zag foam (4.25 GPa).
The shear moduli have a similar tendency and the smallest value was found for the (7,7) armchair system (0.1 GPa).
The most stiff carbon foam is the (1,1) zig-zag with $\textbf{B}=285.13$ GPa and $\textbf{G}=176.95$ GPa, because this structure is closest to the diamond structure.
On the other hand, it is interesting to note that for a given pore size the armchair carbon foams seem to be mechanically more stable than the zig-zag structures.
Another tendency is that the bulk as well as the shear moduli become smaller going from systems with $M=1$ to $M=5$.
\begin{table}
\caption{The bulk and the shear moduli (\textbf{B} and \textbf{G}) of zig-zag carbon foams given in GPa.\cite{Kuc2006}}
\label{tab:3}
\begin{tabular}{p{0.8cm}p{0.95cm}p{0.95cm}p{0.95cm}p{0.95cm}p{0.95cm}|p{0.95cm}p{0.95cm}p{0.95cm}p{0.95cm}p{0.95cm}}
\hline\noalign{\smallskip}
& \multicolumn{5}{c|}{\textbf{B}} & \multicolumn{5}{c}{\textbf{G}} \\
\hline\noalign{\smallskip}
\textbf{\backslashbox {$N$}{$M$}} & \textbf{1} & \textbf{2} & \textbf{3} & \textbf{4} & \textbf{5} & \textbf{1} & \textbf{2} & \textbf{3} & \textbf{4} & \textbf{5} \\
\hline\noalign{\smallskip}
\textbf{1} & 285.1 & 89.0 & 21.5  & 8.25 & 4.25 & 176.95 & 53.1 & 15.3 & 4.3 & 3.6 \\
\textbf{2} & 265.5 & 157.7 & 54.85 & 18.3 & 6.95 & 105.8 & 32.3 & 13.5 & 6.6 & 3.2 \\
\textbf{3} & 225.2 & 172.9 & 97.0 & 69.5 & 21.0 & 122.7 & 17.9 & 8.7 & 8.1 & 4.6 \\
\textbf{4} & 148.6 & 137.3 & 124.8 & 73.9 & 69.1 &  142.8 & 11.6 & 6.3 & 4.6 & 4.6 \\
\textbf{5} & 183.5 & 107.9 & 109.4 & 98.9 & 75.5 &  76.5 & 7.95 & 5.0 & 3.8 & 2.9 \\
\hline\noalign{\smallskip}
\end{tabular}
\end{table}

\begin{table}
\caption{The bulk and the shear moduli (\textbf{B} and \textbf{G}) of armchair carbon foams given in GPa.\cite{Kuc2006}}
\label{tab:4}
\begin{tabular}{p{0.8cm}p{0.95cm}p{0.95cm}p{0.95cm}p{0.95cm}|p{0.95cm}p{0.95cm}p{0.95cm}p{0.95cm}}
\hline\noalign{\smallskip}
& \multicolumn{4}{c|}{\textbf{B}} & \multicolumn{4}{c}{\textbf{G}} \\
\hline\noalign{\smallskip}
\textbf{\backslashbox {$N$}{$M$}} & \textbf{2} & \textbf{3} & \textbf{4} & \textbf{5} & \textbf{2} & \textbf{3} & \textbf{4} & \textbf{5} \\
\hline\noalign{\smallskip}
\textbf{1} & 266.7 & 78.2  & 28.8  & 12.6 & 64.35 & 24.65 & 12.45 & 7.9 \\
\textbf{2} & 213.3 & 156.3 & 57.4  & 23.4 & 26.2  & 16.8  & 9.2   & 6.4 \\
\textbf{3} & 140.6 & 162.1 & 95.7  & 47.2 & 12.2  & 8.7   & 4.7   & 4.3 \\
\textbf{4} & 89.6  & 114.8 &       &      & 9.0   & 4.3   &       &     \\
\textbf{5} & 65.0  & 88.2  & 100.6 & 92.3 & 3.6   & 3.7   & 2.4   & 0.9 \\
\hline\noalign{\smallskip}
\end{tabular}
\end{table}

As expected for structures built from graphite and diamond segments containing both \textit{sp}$^2$ and \textit{sp}$^3$ hybridized atoms, the calculated bulk moduli of the carbon foams vary over a wide range: $\sim$5 up to $\sim$300 GPa, i.e. ranging between that of graphite (5.5 GPa) and nearly approaching that of diamond (442~GPa\cite{McSkimin1972}).
Moreover, the (3,1) armchair and SDG carbon foams also possess rather large bulk moduli of 48.5 GPa and 20.2 GPa, respectively, compared to graphite.

The \textbf{G} values of carbon foams are, however, clearly smaller than those of diamond (621 GPa) and for larger structures they become close to that of graphite (3 GPa).
Evidently carbon foams are mechanically rather stable concerning the bulk moduli.
However, it is important to notice that larger foams [Tab.~\ref{tab:3} and Tab.~\ref{tab:4}] could become unstable against shear forces, because of their smallshear moduli.
This behavior comes from the fact that carbon foams are highly anisotropic systems.

Furthermore, the properties of the investigated carbon foams with $M=2$ and increasing $N$ were studied to search for size dependent trends.
The results are shown in Table~\ref{tab:5}.
Indeed, these carbon foams slowly approach the properties of layered graphite.
The bulk modulus reaches the maximum value at the (3,2) structure and decreases continuously with increasing sizes.
As the foam with increasing size starts to mimic the structure of layered graphite, its binding energy increases as well.
However, the mass densities are much smaller, because the distance between graphitic segments is over two times larger than in layered graphite.
\begin{table}
\caption{Calculated mass densities ($\rho$), bulk moduli (\textbf{B}), band gaps ($\Delta$) and binding energies per atom (E$_{\rm bind}$) of zig-zag carbon foams with $M=2$.}
\label{tab:5}
\begin{tabular}{p{2.0cm}p{2.0cm}p{2.0cm}p{2.0cm}p{2.0cm}}
\hline\noalign{\smallskip}
\textbf{({$N$},{$M$})} & {$\rho$ (g~cm$^{-3}$)} & \textbf{B} (GPa) & {$\Delta$ (eV)}  & {E$_{\rm bind}$ (eV~atom$^{-1}$)} \\
\hline\noalign{\smallskip}
\textbf{(1,2)}  & 1.77 & 89.0  & 1.48 & 7.745 \\
\textbf{(2,2)}  & 1.46 & 157.7 & 1.56 & 7.796 \\
\textbf{(3,2)}  & 1.30 & 172.9 & 0.0  & 7.817 \\
\textbf{(4,2)}  & 1.20 & 137.3 & 1.11 & 7.836 \\
\textbf{(5,2)}  & 1.14 & 107.9 & 0.78 & 7.852 \\
\textbf{(6,2)}  & 1.09 & 84.1  & 0.0  & 7.863 \\
\textbf{(7,2)}  & 1.06 & 74.2  & 0.63 & 7.873 \\
\textbf{(8,2)}  & 1.03 & 66.6  & 0.50 & 7.881 \\
\textbf{(9,2)}  & 1.01 & 58.8  & 0.0  & 7.887 \\
\textbf{(10,2)} & 0.99 & 51.4  & 0.44 & 7.893 \\
\textbf{graphite} & 2.27 & 5.5  & 0.0 & 7.986 \\
\hline\noalign{\smallskip}
\end{tabular}
\end{table}

\section{Electronic Properties}
\label{subsec:4}

In this chapter the densities of states (DOS) and band structures of the investigated carbon foams are discussed.
It is interesting to point out, that the calculated band gaps of zig-zag foams indicate a similar size dependence as for zig-zag carbon nano\-tubes.
They are metallic, if the distance between two junctions is a multiple of three hexagonal units:
\begin{equation}
\label{eq:9}
(N,M)=[3\times m,M]
\end{equation}
and/or
\begin{equation}
\label{eq:10}
(N,M)=[N,3\times m]
\end{equation}
with $m=1, 2, 3,\ldots$; otherwise the foams are semiconducting with a gap size in similar range as for semiconducting carbon nano\-tubes (0.6$-$1.0~eV).
Similar to armchair carbon nano\-tubes, the armchair carbon foams are all metallic independent of their size.

\begin{figure}[ht]
\includegraphics[scale=.65]{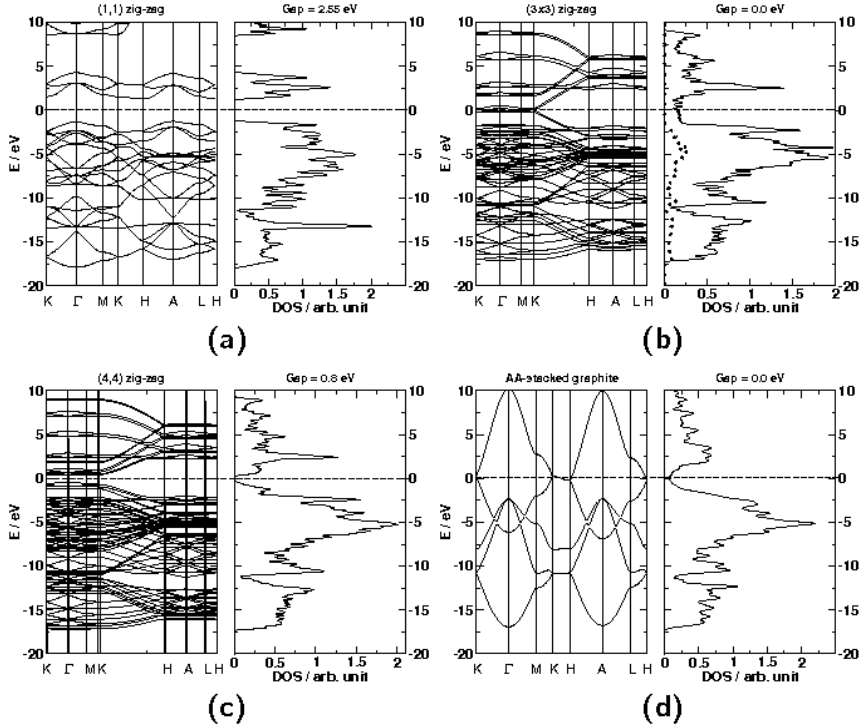}
\caption{The band structures and the densities of states of the symmetric zig-zag carbon foams (a$-$c) and graphite (d) in a hexagonal lattice representation. The dotted line in (b) denotes the PDOS of the \textit{sp}$^3$ carbon atoms along the junction. The Fermi level is shifted to 0.0~eV (horizontal dashed lines).\cite{Kuc2006}}
\label{fig:9}
\end{figure}

Figure~\ref{fig:9} shows the band structures and densities of states for the symmetric zig-zag carbon foams calculated in hexagonal lattices.
These results are compared with the band structure and DOS of AA-stacked hexagonal graphite.
Although the dispersion along the lowest conduction band as well as  the highest valence band is very small, the systems were recognized as indirect-gap semiconductors.
A distinct dispersion appears along the K$-$H lines [Fig.~\ref{fig:9}b and Fig.~\ref{fig:9}c].
The electronic structure reveals that mostly the atoms in the direct vicinity of the \textit{sp}$^3$ carbon chains contribute to the bands near the Fermi level.
The (3,3) zig-zag carbon foam is metallic with bands crossing the Fermi level at the K point of the Brillouin zone.

\begin{figure}[ht]
\includegraphics[scale=.65]{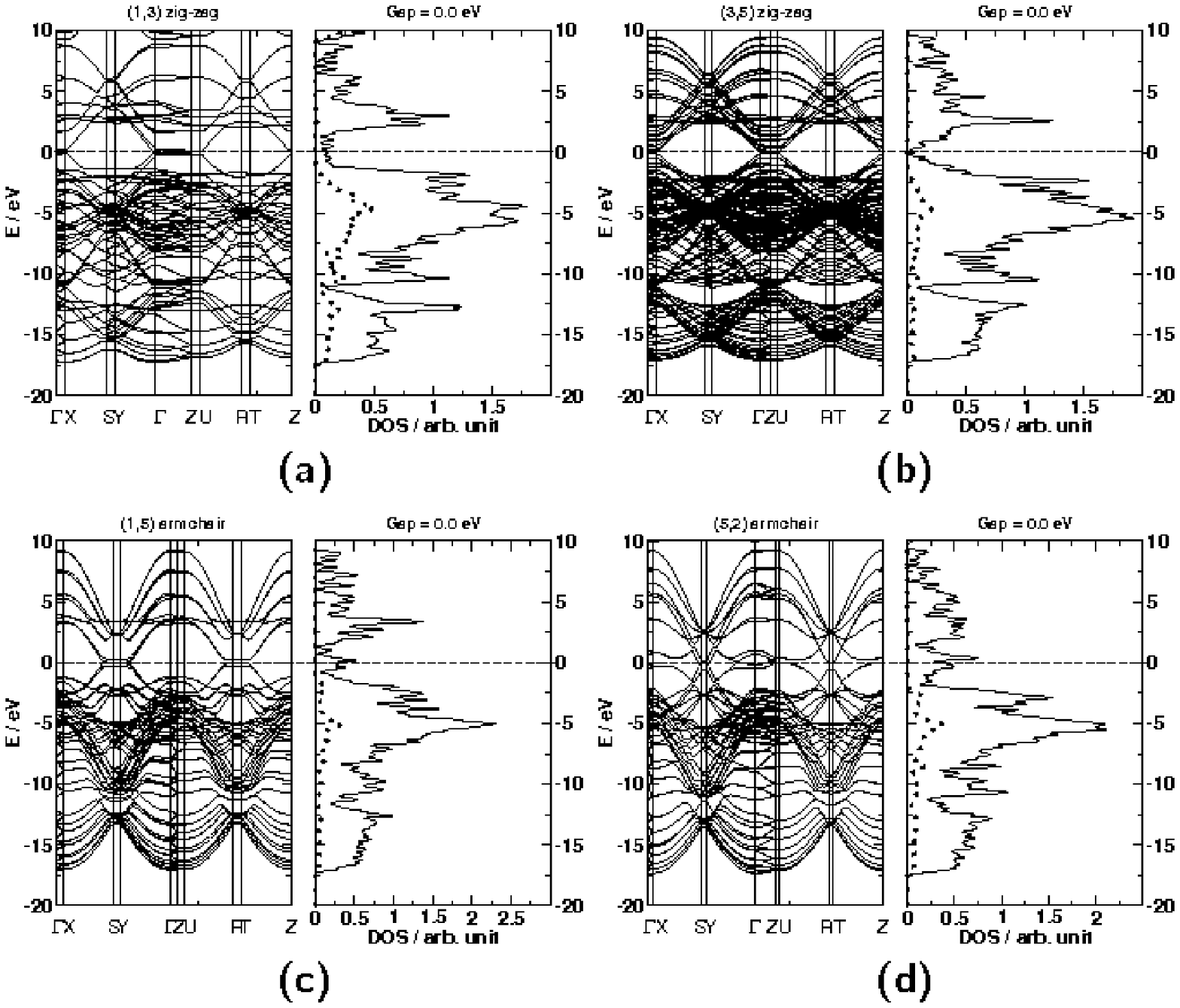}
\caption{The band structures and the densities of states of zig-zag (a, b) and armchair (c, d) carbon foams in an orthorhombic lattice representation. The dotted lines denote the PDOS of the \textit{sp}$^3$ carbon atoms along the junction. The Fermi level is shifted to 0.0~eV (horizontal dashed lines).\cite{Kuc2006}}
\label{fig:10}
\end{figure}

Some examples of band structures and DOS of the orthorhombic ($N\neq M$) zig-zag carbon foams are shown in Fig.~\ref{fig:10}a and Fig.~\ref{fig:10}b.
These foams are metallic as it was stated in Eq.~\ref{eq:9} and Eq.~\ref{eq:10}.
There is a visible large band dispersion along the $k_a-k_b$ plane [see Fig.~\ref{fig:3} for definition of $k_a$ and $k_b$], similar to that of a graphene monolayer.

Band structures and DOS of some armchair carbon foams are shown in Fig.~\ref{fig:10}c and Fig.~\ref{fig:10}d.
This group of metallic structures has bands crossing the Fermi level along the X$-$S and Y$-\rm{\Gamma}$ lines.
Large dispersions of valence and conduction bands are visible as well.

PDOS of the junction atoms (\textit{sp}$^3$) is highlighted in Fig.~\ref{fig:9} and Fig.~\ref{fig:10} for the metallic systems.
It can be seen that the line of junction atoms has an insulating character and the metallic properties of the carbon foams are restricted to the graphene-like stripes with \textit{sp}$^2$ hybridized carbon atoms.
Thus, for larger semiconducting carbon foams the band gap will decrease with increasing size [cf.\ Table~\ref{tab:5}].

\begin{figure}[ht]
\includegraphics[scale=.40]{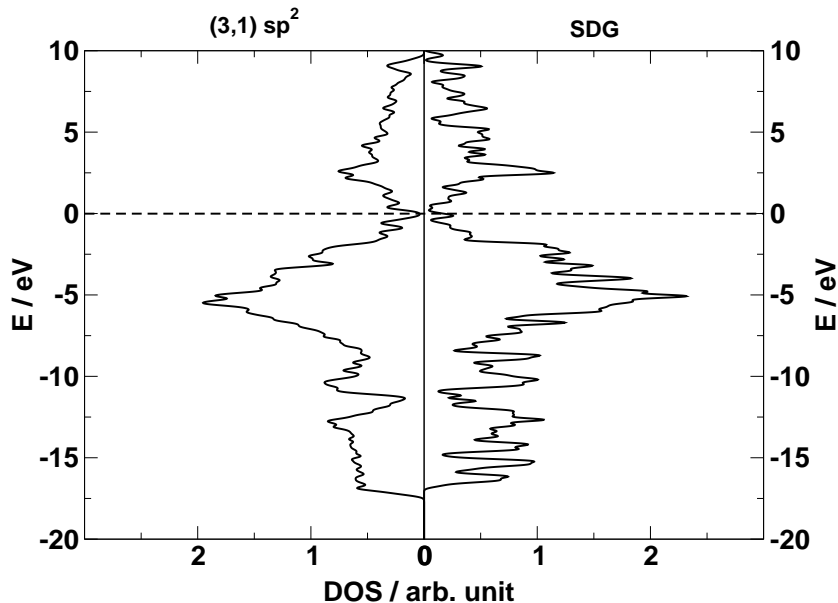}
\caption{The densities of states of the (3,1) armchair carbon foam (left) and the screw dislocated graphite (right). The Fermi level is shifted to 0.0~eV (horizontal dashed line).}
\label{fig:11}
\end{figure}

The family of \textit{sp}$^2$ armchair foams is also metallic.
The densities of states of the (3,1) structure and the SDG are shown in Fig.~\ref{fig:11}.
The band structures are very complicated and therefore not shown here.
The DOS of both systems are similar to that of layered graphite.

\section{Defected Graphite}
\label{subsec:5}

\begin{figure}[ht]
\includegraphics[scale=.72]{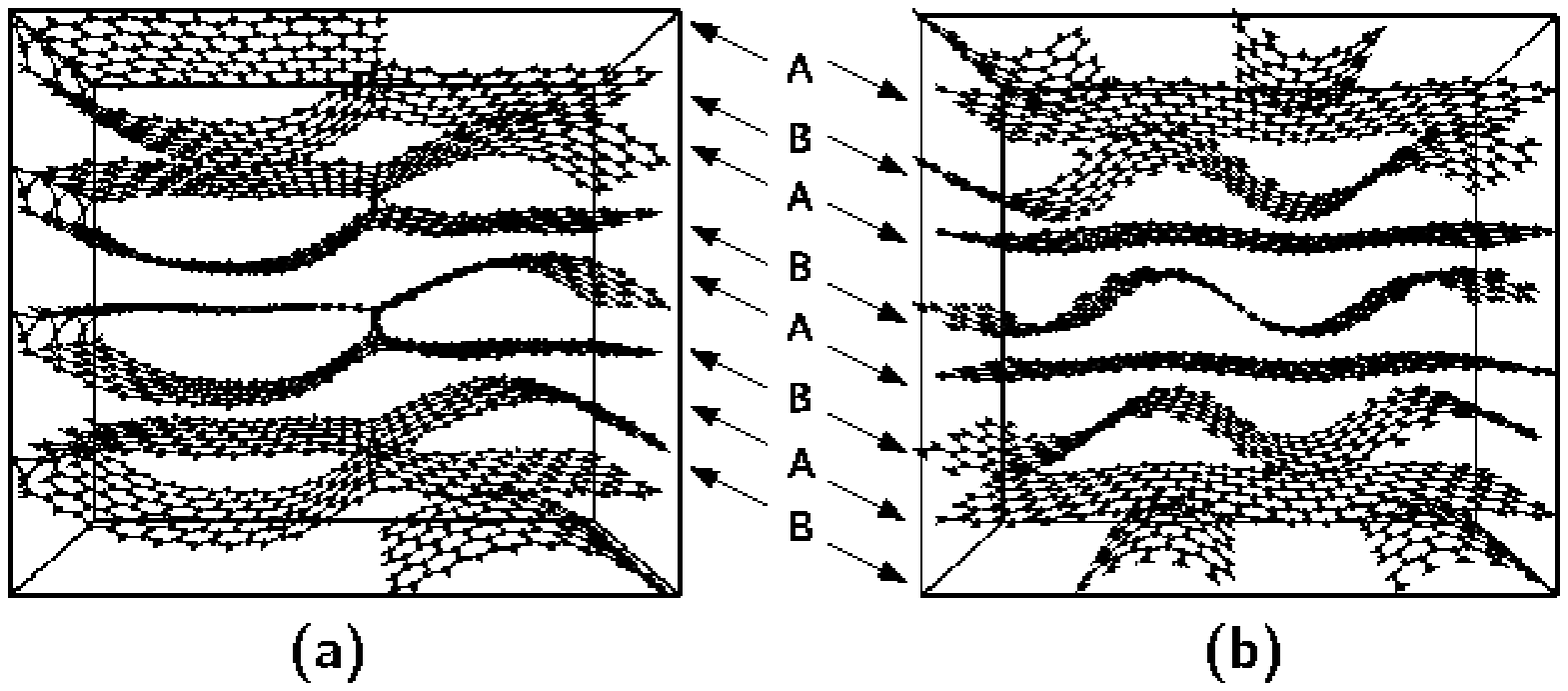}
\caption{(a) The zig-zag and (b) the armchair prismatic edge dislocations in layered graphite. A and B indicate the stacking of layers.}
\label{fig:12}
\end{figure}

In this chapter, two of the recently reported\cite{Suarez2007} dislocations are investigated: the zig-zag and the armchair prismatic edge dislocations.
Under strong irradiation the original layered structure of graphite ($d=3.35$~\AA) is lost.
Two types of highly defected graphite structures are shown in Fig.~\ref{fig:12}.
Both are formed from ABAB-stacked graphite.

In the zig-zag prismatic edge dislocated system the neighboring layers are interconnected locally by \textit{sp}$^3$ carbon atoms [Fig.~\ref{fig:12}a].
This structure is similar to the carbon foams discussed above.
Unlike in the carbon foams, the \textit{sp}$^3$ carbons connect only two graphitic stripes with \textit{sp}$^2-$\textit{sp}$^3$ C bonds of 1.491~\AA.
The occurrence of \textit{sp}$^3$ hybridized atoms causes a reorientation of the single layers leading to the formation of connected double layers (containing loops or cavities).
The cavities have a width of 6.7~\AA.
Each double layer is repeated with about 10~\AA\ distance, thus the minimum distance between two graphitic fragments is as large as in layered graphite (3.3~\AA).

On the other hand, the armchair prismatic edge dislocations are characterized by formation of wrinkled not bonded graphene layers [Fig.~\ref{fig:12}b].
In this case, well-defined cavities and a $d$-expansion appear due to the bends of the graphene layers.
No direct connections between the neighboring layers exist and thus, the system is built only from \textit{sp}$^2$ hybridized carbon atoms.
This structure can be considered as an intercalated graphite.
Here the intercalant is simply a curled graphene layer.
The bond lengths are typical for graphite.
The maximal separation between the flat and the curled layers is about 6.8~\AA, whereas the minimum is around 3.2~\AA.
The bent layers can be considered as spacers in the AA-stacked graphite with an interlayer distance of 10.0~\AA.

As both structures have pores with diameters larger than the van der Waals distance in graphite, the mass densities decrease from $\rho=2.27$~g~cm$^{-3}$ (graphite) to $\rho=1.62$~g~cm$^{-3}$ for both dislocated graphite modifications.
They are very stable having binding energies of 7.88~eV~atom$^{-1}$ and 7.94~eV~atom$^{-1}$ for zig-zag and armchair dislocations, respectively.

\section{Glitter}
\label{subsec:6}

\begin{figure}[ht]
\includegraphics[scale=.90]{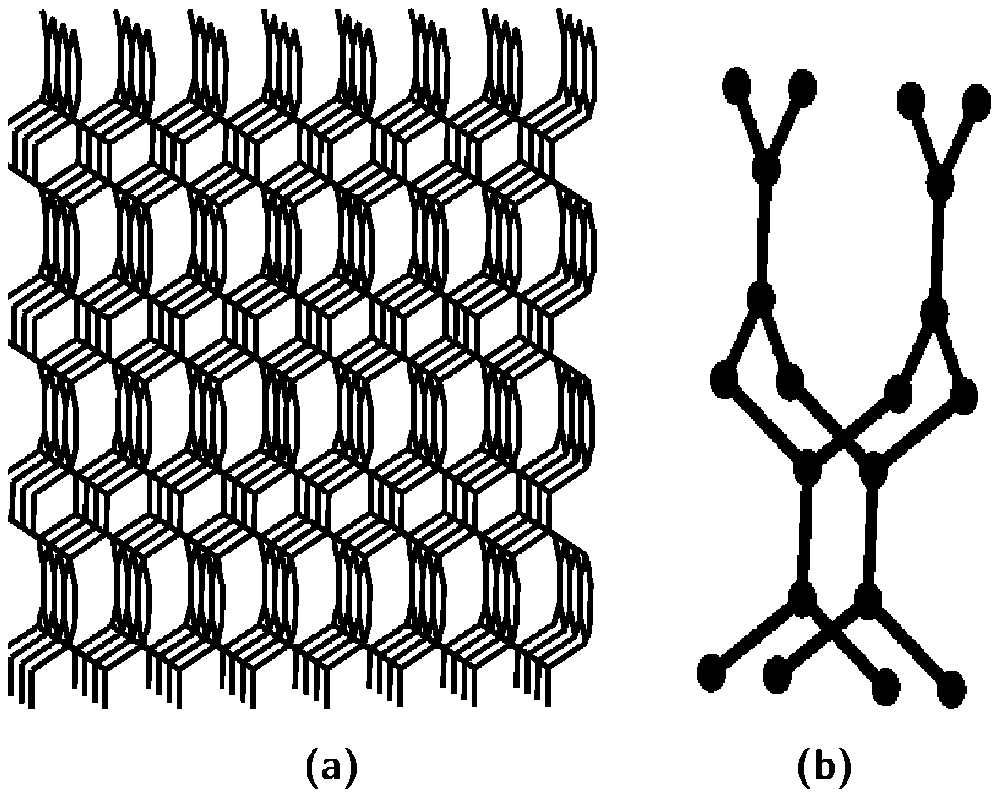}
\caption{An extended view (a) and the unit pattern (b) of glitter.}
\label{fig:13}
\end{figure}
Glitter was first proposed in 1994 by Bucknum and Hoffmann\cite{Bucknum1994} as a potential allotrope of carbon.
The original intent was to combine the archetypal trigonal planar, 3-connected bonding of carbon in graphite with archetypal tetrahedral, 4-connected bonding of carbon in diamond.
Such a material is known as a 3-, 4-connected carbon network and belongs to the \textit{sp}$^2+$\textit{sp}$^3$ group.
Glitter is a hypothetical structure constructed from a structural basis constituted by a 1,4-cyclohexadienoid motif.
The extended structure of glitter together with its unit pattern is shown in Fig.~\ref{fig:13}.
This system resemble the carbon nanofoam structure but with 3D interconnected channels.
In some sense, glitter represents intermediate carbon form between graphite and diamond, similar to small size carbon foams.

In Table~\ref{tab:6} the results from DFTB calculations are presented.
The results agree very well with the previously published DFT calculations of glitter.\cite{Bucknum1994, Bucknum2005, Bucknum2006a}
Mass density of this tetragonal structure is 3.00~g~cm$^{-3}$, intermediate between that of graphite (2.27~g~cm$^{-3}$) and diamond (3.55~g~cm$^{-3}$).
The C--C bonds correspond to the single (1.534~\AA) and double (1.348~\AA) carbon bond lengths.
The bulk modulus calculated at the DFTB level here is compared to the number obtained from Cohen's semiempirical formula.\cite{Bucknum1994}
This number suggest that glitter is almost as stable as diamond (experimental 442~GPa\cite{McSkimin1972}).
Glitter, with its binding energy of7.52~eV~atom$^{-1}$, is by 0.47~eV~atom$^{-1}$ less stable than graphite (7.99~eV~atom$^{-1}$).
At the DFT calculations\cite{Bucknum1994, Bucknum2005, Bucknum2006a} the authors show that this difference is also around 0.5~eV~atom$^{-1}$.
\begin{table}
\caption{Calculated unit cell parameters (a, b, c), mass densities ($\rho$), C--C bond lengths, and bulk moduli (\textbf{B}) of glitter.}
\label{tab:6}
\begin{tabular}{p{2.5cm}p{2.5cm}p{2.5cm}}
\hline\noalign{\smallskip}
\textbf{property} & \textbf{DFTB} & \textbf{DFT}\cite{Bucknum1994, Bucknum2005, Bucknum2006a} \\
\hline\noalign{\smallskip}
\textbf{a=b}                              & 2.564~\AA & 2.560~\AA \\
\textbf{c}                                & 6.064~\AA & 5.925~\AA \\
\textbf{$\rho$}                             & 3.00~g~cm$^{-3}$  & 3.12~g~cm$^{-3}$  \\
\textbf{\textit{sp}$^2-$\textit{sp}$^2$}  & 1.348~\AA & 1.350~\AA \\
\textbf{\textit{sp}$^2-$\textit{sp}$^3$}  & 1.534~\AA & 1.510~\AA \\
\textbf{B}                                & 397~GPa   & 440~GPa   \\
\hline\noalign{\smallskip}
\end{tabular}
\end{table}

\begin{figure}[ht]
\includegraphics[scale=.45]{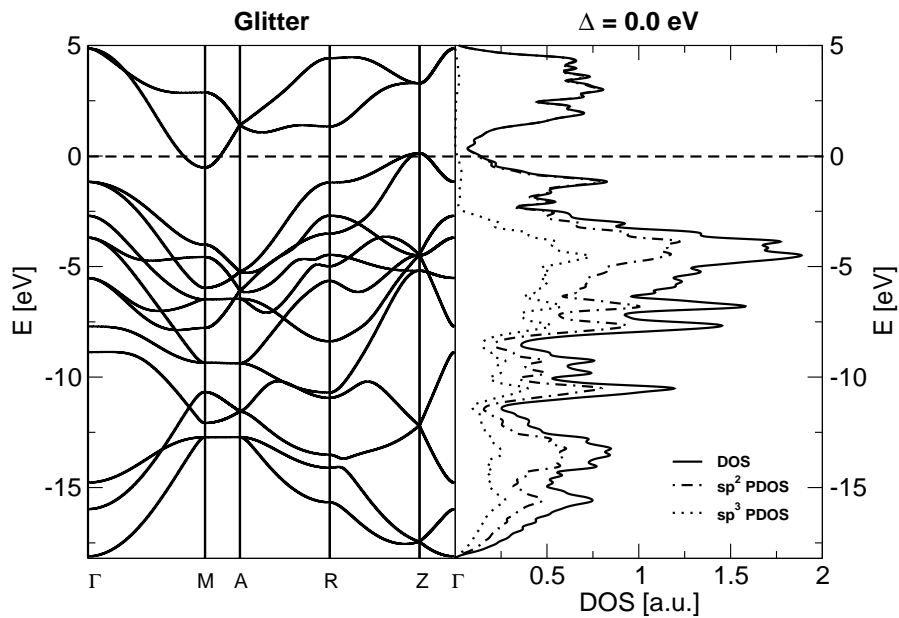}
\caption{The calculated band structure (left) and density of states (right) of glitter. The black, green and red lines denotes the total DOS, partial DOS ofthe \textit{sp}$^3$ and \textit{sp}$^2$ carbon atoms, respectively. The Fermi level is shifted to 0.0~eV (horizontal dashed line).}
\label{fig:14}
\end{figure}

The electronic properties of glitter (band structure and DOS) suggest metallic character of the system.
The same result was obtained at the DFT LDA level.
The band structure and DOS, with highlighted contributions of \textit{sp}$^3$ and \textit{sp}$^2$ carbon atoms, are shown in Fig.~\ref{fig:14}.
It can be seen that the $\pi$* band dips down into the occupied bands of glitter at symmetry point M in the reciprocal space.

\section{Conclusions}
\label{sec:2}

In this work hypothetical carbon allotropes, called carbon nanofoams, have been discussed concerning the stability and the electronic properties.
The construction shows that the chain of the sp$^3$ - hybridized atoms along the junction are connected covalently with layers of graphite stripes having either zig-zag or armchair types of edges.

The results of DFT based computations confirm high stability of carbon foams as compared to the most stable carbon allotropes (graphite and diamond).
These systems may represent novel porous carbon modifications with sp$^2$-sp$^3$ hybridization and high structural stability at low mass density.
The foams have large bulk moduli, although they might become less resistant against shear forces, when the size of the pores is increased.
The stiffness of carbon foams can be improved by intercalation with e.g.\ carbon nano\-tubes.
The electronic properties of the investigated foams are very similar to those of carbon na\-no\-tubes.
Zig-zag carbon foams are metallic only, if one of the walls has a size, which is a multiple of three hexagonal units.
Otherwise, the zig-zag foams are semiconducting.
Armchair foams have metallic character independent of their pore size.

The investigations on carbon foams are still in progress but the results should encourage experimental investigations for the synthesis of such new carbon systems.
High porosity, crystal-like structure and low mass density are very attractive features, in particular for guest-host interactions.

\section{Acknowledgement}

Support of this research is acknowledged to Stiftung Energieforschung BW. The authors thank also M. Heggie and I. Suarez-Martinez for data about screw dislocated graphite. Figures were made using GTK Display Interface for Structures 0.89.

%\input{referenc_nanofoams}
%\bibliography{refnanofoams.bib}
%\bibliographystyle{spmpsci}
\end{document}